\documentclass[aps,prl,onecolumn,12pt,notitlepage]{revtex4-2}
\usepackage{graphicx}% Include figure files
\usepackage{dcolumn}% Align table columns on decimal point
\usepackage{bm}% bold math
\usepackage{graphicx}
\usepackage{tikz}
\usepackage{tabularx}
\usepackage{array}

\begin{document}

%\preprint{APS/123-QED}

\title{Rethinking Discovery Research with AI: The promise of transdisciplinarity and the danger of reductionism}
\author{Iulia Georgescu\footnote{Institute of Physics. Writing in a personal capacity.}}
%\affiliation{Institute of Physics}%Lines break automatically or can be forced with \\
\author{Venkatesh Narayanamurti\footnote{Emeritus Professor of Technology and Public Policy, Engineering and Applied Sciences and Physics at Harvard University.}}
% \affiliation{Institute of Physics}%Lines break automatically or can be forced with \\

\begin{abstract}
\textit{In theory, transdisciplinary research focuses on solving real-world problems without being constrained by disciplinary boundaries and methodologies. In practice, doing so is challenging, as recognized by the phrase ``The world has problems; universities have departments''. As a powerful integrative tool, AI may help overcome traditional reductionist approaches and enable genuine transdisciplinarity. In this essay, we discuss how this could happen and why, within the current system of incentives, it is not happening.} 
\end{abstract}

\maketitle

AI technologies transform the way we do science, but not just for the reasons one might first think of. AI is fundamentally an integrative tool that can help us break away from the traditional reductionist approach to discovery research and enable genuine transdisciplinarity, which in turn is key to tackling some of the complex societal problems. While it is tempting to think so, in fact the opposite is happening. It is a critical time to rethink the system of incentives that underpin discovery research to ensure that AI improves, not impairs, science and the quest for understanding.

\section{Problems or disciplines?}

In 1952, philosopher Karl Popper noted that ``disciplines are distinguished partly for historical reasons and reasons of administrative convenience (such as the organisation of teaching and appointments), and partly because the theories which we construct to solve our problems have a tendency to grow into unified systems.'' \cite{Popper}

In fact, the historical reasons and administrative convenience are related. Up to the 19th century science did not have clear divisions. Within natural sciences there was little distinction between physics and chemistry, no distinction between fundamental and applied science or among contributors, between amateur and professional scientists \cite{Gross}. As the corpus of knowledge increased, it became necessary to divide it into manageable chunks, which led to specialization. For example, if by mid-18th century the total number of scientific articles per year was under or around 100 (which one could comfortably read), in 1800 there were about 900 articles published per year (already a lot to keep up with). 

Other factors contributed to the transition from natural philosophy to modern science. The professionalization of science pushed amateur scientists to the fringe. Universities reorganised curricula and chairs to reflect these changes, leading to new administrative units within the reformed German universities of the first half of the nineteenth century \cite{Stichweh}. The model spread quickly in Europe and beyond. Meanwhile, academic journals specialised to cater for the different fields and communities crystallized around these areas. 

The separation between pure/basic and applied appeared with the industrial revolution. In 1883, American physicist Henry Augustus Rowland gave a speech called `A plea for pure science’ in which he distinguished between the science of physics and its applications like ``telegraphs, electric lights and such conveniences'' \cite{Rowland}. This dichotomy become enshrined after WWII, due to Vannevar Bush’s 1945 report `Science, the Endless Frontier’ \cite{Bush} despite Bush having cautioned that ``the distinction between applied and pure research is not a hard and fast one''. 

Neither the emergence of scientific disciplines, nor the basic/applied categorisation solved the problem of the ever-increasing amount of knowledge. Quite the contrary, the amount of specialised scientific knowledge kept increasing at pace. ``The marvellously rapid progress of chemistry during the last twenty years has made it difficult for the most industrious cultivator of the science to keep abreast of the knowledge of the day,'' chemist Justus von Liebig noted in 1889 \cite{Gross}. The same words would have resonated with researches in the 1960s and strongly resonate today. Indeed in 1850, over 12,000 articles were published, a number that increased to over 47,000 in 1900. By 1950 (over 200,000 articles published that year) it was already impossible to keep up with publications in one discipline, and many sub-disciplines had already emerged. In 2000, over 1.5 million articles were published and in 2025 6.7 million (Data from Dimensions, excludes preprints and conference proceedings). 

There are many ways to cut the cake, and correspondingly various taxonomies, but for example, the Australian and New Zealand Standard Research Classification ANZSRC (which is used by Dimensions) lists around 2,000 subfields. Take physics for example, subfields include broad areas such as nuclear physics or particle physics that could be further subdivided into research areas. Specialization runs so deep that there are, say particle physicists, who have little familiarity with what other particle physicists work on. These disciplinary boundaries are artificial and become fuzzy at the interfaces. In fact, even for an old discipline such as physics it is quite difficult to draw clear demarcations and an accurate map of the field \cite{Sinatra}.  

So, do we actually need disciplines other than for reasons of tradition and administrative convenience (including information management)? Popper had observed ``But all this classification and distinction is a comparatively unimportant and superficial affair. We are not students of some subject matter but students of problems. And problems may cut right across the borders of any subject matter or discipline.'' \cite{Popper} Another way of putting it is the idea of transdisciplinarity, research that transcends disciplinary boundaries and methodologies and is driven by the need to solve real-world problems \cite{Hadorn}. As appealing as Popper’s view is, it comes in stark contradiction to the reality of the research ecosystem aptly summarized by the phrase ``The world has problems; universities have departments.'' \cite{Kellogg}

Can we become students of problems in today’s complex ecosystem whose incentives reinforce disciplinary silos and strengthen the dichotomy between basic and applied research? Before attempting to answer this question, we will discuss the example of an organization that lived by Popper’s ideal and had an extraordinary legacy in science and technology.

\section{The Bell Labs model}

The AT\&T Bell Labs made great contributions to science and technology over the past 100 years: from inventions that shaped the world we live in today (transistors, lasers, photovoltaic cells and the Unix operating system) to opening new fields such as radio astronomy or information theory. Much has been written on how this was possible \cite{NO2016,Georgescu,NT2024}, but a couple of things stand out. Bell Labs practiced an irreverence for boundaries between disciplines and between so-called `basic’ and `applied’ research. Teams would often consist of researchers without specific credentials in that research area, and that very diversity of approaches and backgrounds would enable breakthroughs. For example, the challenge to develop a reliable solid-state amplifier (the transistor) involved new theoretical physics insights and advances in materials science and engineering for the control of electrons and amplification of signals \cite{NT2021}.

Although the Labs did have divisions (for administrative convenience), they were deliberately organised in a way that these would not create silos. Physically, this was done through the architecture of the buildings at Murray Hill: long corridors with open door offices made encounters unavoidable and the alternation of offices with experimental labs such that theorists, experimentalists and technicians would find themselves in the same space. Psychologically, this was achieved through the egalitarian meritocracy where everyone was equally respected for their ideas and contributions regardless of the background and education credentials.

At Bell Labs, new starters would be given purposely vague direction and encouraged to follow their own interests. Such was the case of John Hopfield, who would later recall in his Nobel Prize Lecture, how his research career at Bell Labs started by looking for a good question, which took him to solid-state physics and from there to biology. ``I was now looking for a big problem, whose resolution and understanding will be of significance far beyond its nominal discipline boundaries, and will reorganize the fields from which they came.'' \cite{Hopfield} The luxury of long-term thinking that Bell Labs researchers could afford meant that they were often thinking of `problems further down the road’ (or `the next-adjacent possible’ \cite{NT2021}) which by their very nature are beyond existing disciplinary boundaries. Transdisciplinarity was therefore deeply embedded in both the ways they worked and the problems they sought.

\section{An opportunity}
While recreating Bell Labs would be challenging today \cite{Georgescu}, there may be another way to break disciplinary boundaries and become students of problems. That is by using the technology that Hopfield’s work help develop: AI.

Going back to the reason why disciplines first emerged, now for the first time, there seems to be a way of keeping abreast of the ever-increasing volume of scientific literature. The ability of large language models to search through the corpus of knowledge and retrieve information suggests that one could use them to gain unprecedented breadth (terms and conditions apply, see following section) across domains. From a philosophical perspective scientific instruments are epistemic enhancers providing extensions to humans’ limited senses. Telescopes and microscopes allow us to see the far and the small, while other instruments let us to see the unseen, such as detecting radiation or magnetic fields. Like computer simulations, AI technologies can be used for extrapolation, conversion and augmentation, but they also provide new enhancer:  integration, synthesizing across large amounts of very different types of information.

As an integrative tool, AI seems to offer a route to transdisciplinarity . Furthermore, AI is found at different interfaces.

\subsection{Between disciplines}
AI as a field of research has always been at the confluence of different disciplines. Think of the participants to the 1956 Dartmouth workshop which is seen as the founding moment for artificial intelligence. They were not yet computer scientists (computer science was just starting to emerge) but mathematicians, electrical engineers, physicists, psychologists and economists. Subsequent developments in the field kept coming from these areas and others. So even if disciplinary boundaries place AI as a sub-discipline of computer science, this is to a degree arbitrary, and there are numerous examples of influential works found at the fuzzy boundaries between disciplines.

Hopfield’s 1982 article `Neural networks and physical systems with emergent collective computational abilities' \cite{Hopfield82}, which is a milestone result in the development of AI and led to the 2024 Nobel prize in Physics, is such an example. The article was classified as biophysics, perhaps at odds with the keywords: associative memory/parallel processing/categorization/content-addressable memory/fail-soft devices. Hopfield’s affiliation was the Division of Chemistry and Biology at Caltech and Bell Labs. Out of the article's cited references 8 are from neuroscience, 3 are multidisciplinary, 3 biological mathematics, 2 psychology, 2 biological science, 2 physical sciences, 1 theory of computation. This a truly interdisciplinary paper (and has been identified as such in the analysis in \cite{Sinatra}) that illustrates the superficiality of classifications and distinctions.

\subsection{Between basic and applied}
On the matter of classifications, one would likely assign AI technologies to engineering research. However, the development of these methods benefits from more foundational research and feeds into the further evolution of science and technology. An example of the former are diffusion models, enabling image generation. These models were inspired by non-equilibrium statistical physics. As for the latter, the theory of neural networks turns out to be useful in very pure areas of research in physics such as string theory.

\subsection{A tool and a system to study}
Condensed matter physics and materials physics is possibly the largest sub-discipline of physics today, but until the 1950s it was considered fringe, even distasteful (Wolfgang Pauli called it “dirty physics”). The reason for its success is that it provided an arena to develop new active tools and devices (think of all the applications of new materials), which in turn allowed the observation of new phenomena (for example high-temperature superconductivity, the fractional quantum Hall effect and other emergent collective phenomena). Similarly, AI is a tool that provides new capabilities (think of AlphaFold), but at the same time it is a fascinating complex system to be studied. Situations where the tool is also the object of research are relatively rare, and if we are to generalise, very fruitful for both scientific and technological advances.

Considering all the above, it is tempting to conclude that AI technologies present an opportunity to achieve true transdisciplinary and do research in an integrated way focusing on problems. But as is the case with any technology, it’s important to develop and carefully read the fine print.

\section{The risks}
There are various risks related to the use of AI tools in scientific research. On the one hand, there are the problems of inaccuracy, bias, plagiarism and deliberate misinformation. On the other hand, there are concerns associated with narrowing down of research topics and emergence of scientific monocultures, the “illusion of explanatory breadth” \cite{Messeri}, loss of serendipity, deep understanding and creativity. We do not discuss these here (see \cite{Bianchini} for a review), but rather add another risk: exacerbating reductionism.  

Instead of a nurturing a transdisciplinary approach we already see the seeds of reductionism sprouting. AI for science (meaning `AI tools in discovery research and scientific approaches to study AI’) is already breaking up in `AI for chemistry’, `AI for mathematics, etc. This suggests a disciplinary-approach rather than a problem-focused one (AI for drug and material discovery are exceptions). Furthermore, the idea of automated end-to-end science is an extreme form of reductionism, because it involves breaking up research into independent modules whose inputs and outputs can be connected in a linear diagram. Why is this a problem?

Reductionism is a powerful approach. Decomposing complex systems into smaller components that can be described by models and theories, is how a lot of scientific progress has been made. But as effective as reductionism has proved to be it is not almighty. It only goes one way. Reductionism does not imply constructivism, so from the smaller parts one cannot necessarily build the complex system which can behave differently from what would be expected from the sum of individual components (what is known as emergence). 

This simple, yet profound, observation was articulated by Bell labs researcher and Nobel-prize winner, physicist Phil W. Anderson in 1972 ``the ability to reduce everything to simple fundamental laws does not imply the ability to start from those laws and reconstruct the universe'' \cite{Anderson}. Decades later, it is broadly recognized, and often forgotten. For example, the artificial boundaries between disciplines and between basic and applied research are the result of reductionist thinking. 

Both reductionism and constructivism are needed in discovery research to define the problems and to solve them. An illustration of how they work together is search for answers and questions. The technoscientific method developed in \cite{NT2021} proposes that “science draws on technology to discover new facts, while technology draws on science to invent new forms with which to fulfill human desired functions.” From this perspective technoscientific knowledge consists of networks of question-answer pairs that are combined and evolved into new pairs expanding the domain of what is known.

``Answer-finding begins with reductionism, reducing problems into constituent functionalities and analysing whether those constituent functionalities can be met or solved, and ends with integration, integrating the constituent functionalities into a synthesized solution.'' In contrast, ``question-finding, begins with synthesis, taking answers or solutions that already exist and integrating them to discover new questions they might answer or problems they might solve. Then once a question has been tentatively found, it must be followed by reductionist analysis to verify that the answer/solution does indeed answer/solve the question or problem.'' This asymmetry makes question finding more difficult. ``Because question-finding looks upward in the knowledge network, it cannot draw, the way answer-finding can, on a systematic reductionist and analytic hunt for answers deeper within a discipline. The hunt for questions typically lies outside the discipline, hence it depends more on serendipity and chance collision of ideas.''  \cite{NT2021}

Because answer-finding can be tackled with a systematic reductionist and analytic approach it is more favoured within the existing incentive structure than question-finding. Integrative approaches are costly and risky. Furthermore, answers are easier to verify than it is to assess whether questions are good ones (it is often the case that great questions may first appear ludicrous). The problem is exacerbated by the bias towards utility rather than understanding (illustrated by the use of scientific problems as benchmarks for the capabilities of AI models), which again favours the search for answers rather than the formulation of new questions. This is why the use of AI tools in research poses a risk of reinforcing reductionism. 

\section{A route forward}

It’s clear that developments in AI technologies will continue at pace, driven by a complex tangle of incentives, and will bring significant change in how discovery research is done. It’s important to consider carefully what needs to be adjusted at system and individual level so that this change is beneficial to the scientific community and society. We suggested that AI tools, if used appropriately, could contribute to a transdisciplinary approach. However, AI technologies alone are not sufficient to do so.

\subsection{System}
At the system level, we need to balance utility with understanding. To do this it is important to clarify that science is to understand new things, while engineering is to build new things. Discovery and invention are the discrete, and highly desirable, outcomes of science and technology working together. But research in both science and engineering is a continuous, dynamic and complex process, that is more than the sum of its parts. And this process is as important as the outcomes. Narrowly optimizing for outcomes is unlikely to yield the desired results. 

Recent examples illustrate this. AI tools have been used very successfully to predict the structure of new materials with various useful properties. However, this success generated other challenges: how to actually make these materials and how to test the millions of possible compounds? As the work shifts into using AI to tackle the synthesis of new materials, one can imagine further challenges down the line in how to characterize and understand these materials. This AI approach is in stark contrast to the development of the transistor, and subsequently many other materials, where teams of experimentalists, theorists, technicians with backgrounds in physics, chemistry, engineering worked closely together making advances on all fronts. Another area where AI is very successful is mathematics, but recent advances have spurred concern that heavy AI-assisted proofs hide the insights and understanding that the process of finding the proofs uncovers.  Furthermore, while the many goals of mathematics research have been historically positively correlated with one another, excessive optimization for one or two goals may cause them to diverge \cite{Tao}.

These examples show that the process matters as much, if not sometimes more, than the end result. Often getting the final answer is not enough. The process may not be straightforward but is necessary to gain a solid understanding both in experimental and purely-theoretical investigations. These examples are also cautionary tales for the reductionist approach to breaking research down into modules. The risk is that optimizing the parts does not guarantee improvements of the whole (a lesson that is increasingly clear in applying AI in various domains, not only mathematics). 

It’s worth remembering that ``Creativity in science does not result from a static, repetitive, and well-delineated process as found in assembly-line manufacturing. Rather, creative output emerges out of a dynamic, incessantly changing, and ill-defined process'' \cite {Simonton}. Or, as Julio Ottino and Brian Uzzi recently put it, ``The structure of scientific workflows can be treated like a clock. But science as a system of discovery is a cloud. Optimizing each component does not guarantee the integrity of the whole. Optimization may, in fact, undermine it.'' \cite{OU2026}

\subsection{Individual}
At individual level, AI can provide human researchers with more breadth than was ever possible before. However, it cannot directly provide depth, because, most likely, it can only be achieved by doing. The depth is given by active knowledge developed through constant questioning and a lot of hard work (access to tacit knowledge is an important part of building depth). Neither breadth nor depth alone enable transdisciplinarity. Both are needed ``because of the requirement of depth to enable maximum impact of breadth there seems to be a `sweet spot’ in knowledge breadth for maximizing creativity'' \cite{NT2021}. 

We can think of breadth as a network of concepts. Operating in this space involves creating new connections, recombining ideas. Depth is crystalized knowledge and operations in this dimension include the creation/use of representations/patterns to reconfigure ideas so they will fit a new combination. However, neither the breadth nor the depth is completely static: the network keeps adding connections, the depth adapts and refines the representations (in the physics metaphor these are different crystalline structures). It is tempting to suspect that breadth is more general, whereas the depth is more particular to the way of thinking of the individual researcher (which would be consistent with the fact that scientists working the same topic can have very different ways of thinking).

Hopfield’s work is a good example of balancing depth and breadth to achieve that sweet spot of scientific creativity. But his story also suggests that there is also another important ingredient: the quest for good questions. We refer to this question-seeking mindset as being an informed contrarian. It means pushing the limits of understanding by looking at things in a different way, questioning what is established, thinking of new directions beyond what is known. The informed contrarian relies on breadth and depth to glimpse into the `next-adjacent possible’ \cite{NT2021}.

We saw that AI tools could help achieve more breadth (recent examples show how that can be used to tackle problems in different fields \cite{Zhao}). Perhaps now, more than ever before, there is need for depth of expertise and a critical, questioning mindset, neither of which can be gained without sustained effort. There may be ways in which AI technologies could also contribute to developing depth and cultivating an informed contrarian mindset. However, the risks identified so far, to which we added exacerbating reductionism, seem to point in the opposite direction. What seems clearer is that neither the opportunity AI presents can be exploited, nor the risks it entails tackled, unless the scientific community critically evaluates the role of AI in discovery research and explores how the existing system of incentives needs to adapt.

\bibliography{biblio}% Produces the bibliography via BibTeX.

\end{document}